\documentstyle[preprint,amsfonts,aps]{revtex}
\begin{document}
\renewcommand{\theequation}{\arabic{section}.\arabic{equation}}
\begin{titlepage}
\begin{center}
{\Large\bf  Bosonization, Vicinal Surfaces, and \medskip\\ Hydrodynamic
Fluctuation Theory}
\bigskip\bigskip\bigskip\\
{\large Herbert Spohn}\bigskip\\
Zentrum Mathematik and Physik Department, TU M\"unchen,\\
80290 M\"unchen, Germany\medskip\bigskip\\
email: {\tt spohn@mathematik.tu-muenchen.de}
\end{center}
\vspace{3cm}
\noindent
{\bf Abstract.} Through a Euclidean path integral we establish that the 
density fluctuations of a Fermi fluid in one dimension are related to 
vicinal surfaces and to the stochastic dynamics of particles 
interacting through long range forces with inverse distance decay. In 
the surface picture one easily obtains the Haldane relation and 
identifies the scaling exponents governing the low energy, Luttinger
liquid behavior. For the stochastic particle model we develop a 
hydrodynamic fluctuation theory, through which  in some cases
the large distance
Gaussian fluctuations are proved nonperturbatively.\bigskip
\end{titlepage}
\section{Introduction}\label{a}
\setcounter{equation}{0}
As pointed out by Haldane some time ago \cite{H1,H2}, spinless fermions in
one dimension interacting through a short range potential have
universal ground state correlations. The universal properties are
computed on the basis of the Tomanaga--Luttinger hamiltonian
\cite{T,L,Sch} where low energy characteristics turn out to be labelled
by two free parameters, traditionally denoted as the renormalized
sound velocity $v_s ~\mbox{and}~ K .$ These parameters must be
matched to the microscopic Fermi liquid under consideration. In
fact, we will see that they are given  by suitable second
derivatives of the ground state energy/length.

The two most prominent predictions of the Luttinger liquid scenario
are\\
(i) The momentum distribution behaves as
\begin{equation}\label{a.a}
	\langle a^{\dagger}(k) a(k) \rangle  \simeq 
  |k - k_F|^ \alpha \,
 \mbox{sign}(k - k_{F})+\mbox{regular part}
\end{equation}
close to the Fermi momentum $k_F.$
Compared to the noninteracting case, the Fermi fluid looses its
gap at $k_F$ and the Fermi ``surface'' is retained  as power
law singularity only with anomalous 
exponent $\alpha = \frac{1}{2} (K+ \frac{1}{K}-2).$\\
(ii) The density fluctuations in the ground state are severely
suppressed and strongly correlated. Nevertheless they have
Gaussian statistics.
This is a consequence of the bosonization of the density field,
which is the basic observation leading to the exact solution of
the Tomanaga--Luttinger hamiltonian \cite{MLi}.

The anomalous momentum distribution (\ref{a.a}) has been studied by G.
Gallavotti and coworkers \cite{BeGa} through a rigorous implementation
of a renormalization group zooming onto the Fermi surface. The
present paper focuses on the density fluctuations (ii). We
recall first that for the ideal Fermi fluid the structure function, 
i.e. the Fourier transform of the density-density correlations is 
given by
\begin{eqnarray}\label{a.b}
S(k) = \left\{ \begin{array}{ll}
|k|/2\pi  & \mbox{for} ~  |k| \leq 2k_{F}\, ,\\
\rho  & \mbox{for} ~ |k| \geq 2k_{F}
\end{array}
\right.
\end{eqnarray}	
with  $k_{F} = \pi \rho$ the Fermi momentum and $\rho $ the density.  
The interaction smoothens the cusp at $2k_{F}$, similarly to (\ref{a.a}), as 
has been proved recently by Benfatto and Mastropietro \cite{BM}.
For density fluctuations the behavior near $k=0$ is of interest. Here 
the interaction has a much less spectacular effect. It only changes 
the openening angle of the cusp (and of course modifies the straight 
piece). Here we plan to go beyond the static 2-point function and to 
study the small $k, \omega$ behavior of all $n$-point functions, 
including their frequency behavior, i.e. we plan 
to study the generating functional of the density
field for large space-time distances, at which the Gaussian statistics
should be recovered.

As will be explained in detail below the density fluctuations are most 
conveniently investigated through the path integral for the world 
lines of the fermions. This form leads to two other physical 
interpretations. One may think of the world lines as steps of a vicinal 
surface and use the statistical mechanics of surfaces.
In this picture the Gaussian fluctuations are fairly immediate and 
Haldane's parameters are identified as suitable second derivatives of the 
surface tension. In the second interpretation one regards the fermionic 
world lines as trajectories of particles whose motion is then governed 
by  certain stochastic differential equations. Such stochastic 
particle systems are usually described through a hydrodynamic type 
fluctuation theory. In our case the forces between the particles
decay like the inverse distance and are therefore long ranged. We 
will develop a suitable modification of the standard hydrodynamic fluctuation
theory.  In the framework of 
stochastic particle system, at least for some cases, we prove the 
Gaussian fluctuations 
without going through the perturbative double expansion in $n$ and the 
interaction strength.

Different physical interpretations of the same theoretical model 
lead to alternative approximation 
schemes. Properties which look very deep in one formulation are 
physically obvious in another. We regard it as interesting that 
one-dimensional Fermi 
liquids allow for three distinct physical interpretations and try to explore 
their interconnections.
\section{Basic models and their path integrals}\label{b} 
\setcounter{equation}{0}
	
Let us start with the two
prototypical models.\\
(i) Fermions on a ring $[0,\ell]$ interacting through a short range
potential. The hamiltonian reads
\begin{equation}\label{b.a}
H = - \sum_{j=1}^N \frac{1}{2} \frac{\partial^2 }{\partial x_j^2} +
\frac{1}{2} \sum_{i \not=j=1}^N  V(x_i-x_j)\,   .
\end{equation}
We have set $m=1= \hbar$.  $V$ is a short range potential and $x_j \in$
$[0,\ell]$ with periodic boundary conditions. By the Pauli exclusion
principle,
the ground state wave function, $\psi,$ has to vanish at $\{x_i=x_j\}$
(Dirichlet boundary conditions). For density fluctuations the sign
changes in $\psi$ do not matter. Therefore, equivalently, we may
regard (\ref{b.a}) as the hamiltonian of bosons with hard core.
Formally, this corresponds to adding an infinitely strong
repulsive $\delta$-potential to $V$.\\
(ii) Fermions on the periodic lattice $[1, \ldots, \ell]$. In second
quantized form the hamiltonian reads
\begin{equation}\label{b.b}
H = \sum_{x=1}^\ell  \{ -a_x^{\dagger} a_{x+1} - a_{x+1}^{\dagger}
a_x - \Delta a_x^{\dagger} a_x a_{x+1}^{\dagger} a_{x+1} \}\, ,
\end{equation}
$a_{l+1}=a_1$. For future use, we stated only the particular
case of a nearest neighbor interaction. Through the Jordan--Wigner
transformation (\ref{b.b}) turns into the XXZ  hamiltonian as
\begin{equation}\label{b.c}
H = \sum_{x=1}^\ell  \{-\frac{1}{2} (\sigma_x^1 \sigma_{x+1}^1 +
\sigma_x^2 \sigma_{x+1}^2) - \Delta  \sigma_x^3
\sigma_{x+1}^3 \}
\end{equation}
with periodic boundary conditions ${\vec \sigma}_{\ell+1}
= {\vec \sigma}_1$, where ${\vec \sigma}_x$
are the Pauli spin matrices at site $x$.

Both hamiltonians, (\ref{b.a})
and (\ref{b.c}), generate a path integral for the the statistical weight
of fermionic world lines. For (\ref{b.a}) the free measure, $P_o$, are $N$
independent Brownian motions $x_1(t), \ldots, x_n(t)$. By the Pauli 
exclusion principle 
they are constrained not to cross, i.e.
\begin{equation}\label{b.d}
x_j(t) < x_{j+1} (t) ~\mbox{mod} ~ \ell \quad \mbox{for all} ~t~,
j=1, \ldots, N,
\end{equation}
which ensures the Dirichlet boundary condition at $ \{ x_i=x_j
\}$.
The statistical weight of the world lines is then given by
\begin{equation}\label{b.e}
\frac{1}{Z}  P_0 \exp [- \frac{1}{2} \sum_{i
\not= j=1}^N   \int dt V(x_i(t) - x_j(t))] \chi_{NC}\, ,
\end{equation}
where $Z$ is the normalizing partition function and $\chi_{NC}$ is the 
indicator function of the set defined in
(\ref{b.d}). $\chi_{NC}$ restricts the path integral (\ref{b.e}) to 
non--crossing paths only. In the bosonic language this constraint 
corresponds to an infinitely repulsive $\delta$--interaction.

 The statistical weight generated by (\ref{b.c}) is the
same as in (\ref{b.e}). Only the Brownian motion $x_j(t)$ is to be
replaced by a continuous time random walk on $[1, \ldots, \ell]$ with
jump rate 1 to the right and to the left neighbor. In our example
we picked the particular interaction potential $V(x)= - \Delta$
for $| x | =1$ and  $V(x)=0$ otherwise. 

One
physical picture suggests itself immediately. We can think of the
fermionic world lines as the trajectories of a stochastic particle
system. In fact, in the limit $t \to \infty$ it will be a
stationary Markov process. It is a diffusion process in case (i)
and a jump process in case (ii). The quantum mechanical free Bose field
of density fluctuations corresponds to space-time Gaussian
fluctuations in the stochastic particle system. This is the usual
hydrodynamic fluctuation theory as governed by a linear Langevin
equation. Unfortunately, the non--crossing constraint results in
repulsive long range $(1/x)$-type forces between the
particles. Therefore the standard hydrodynamic theory, developed
for short range forces, does not apply. We will explain in Section
\ref{d} the required modifications which are in fact surprisingly
little.

An alternative physical picture is also well known \cite{NRo,BaKa},
but less immediate. We think of $x_j(t)$ as a step of unit height
for a vicinal surface. More explicitely we introduce a height
function $h(x,t)$ for the  height of a crystal surface relative to
the $x,t$ reference plane. Then
\begin{equation}\label{b.f}
 \frac{\partial}{\partial x} h (x,t) = \sum_{j=1}^N  \delta
 (x-x_j(t))\,  .
\end{equation}
In the $t$--direction the average slope vanishes, whereas in the 
$x$--direction it is given by
 the particle density
$\rho=N/\ell$  . Periodic boundary conditions for the particles
means that the surface is extended by repetition at average slope $N/\ell$
along the $x$--axis.

The stepped surface $h(x,t)$ fluctuates with a statistical weight 
given by (\ref{b.e}). In general,
surface fluctuations are expected to be governed by a free massless
Gaussian field in the infrared limit with a strength determined by
the matrix of second derivatives of the surface free energy. Since it
coincides with the ground state energy of the hamiltonian (\ref{b.a}),
we have a direct way to identify the parameters in the Luttinger
hamiltonian. The Haldane relation \cite{H1} then follows as a simple consequence.

The surface picture also indicates the limitations of the
Luttinger liquid concept. The surface free energy may have cusps
and/or flat pieces. The former case corresponds to a roughening
transition where logarithmic fluctuations are suppressed to order
one fluctuations. The latter case is step bunching. The slope 
``segregates'' into macroscopic regions. The steps
are closer together than expected by naive counting.

To give a brief outline. In Section \ref{c} we develop the surface
picture in more detail. In Section  \ref{d} we explore the hydrodynamic
fluctuation theory for the dynamics of world lines and its
consistency with the surface picture. We argue for the validity of
a linear Langerin equation governing the density fluctuations. The
approximations can be controlled for the Calogero-Sutherland model
and for a general system with short range interaction provided the
two-point function scales.  The necessary computations are provided in
an Appendix.

\section{Surface fluctuations and bosonization}
\setcounter{equation}{0}\label{c}
For the sake of concreteness we first discuss fermions on a
lattice, cf. hamiltonians (\ref{b.b}) and (\ref{b.c}). The relations derived
below are general, however. We use
the $\sigma^3$--representation and it is convenient to set up a
corresponding notation. We define $\eta_{x}= (1+\sigma_{x}^{3})/2$. 
Then $\eta_{x} = 0,1$ and we interpret $\eta_{x} = 1 $ as a
surface step at site $x$. A whole step configuration is denoted by 
$\eta$. Then $H$ of (\ref{b.c}) in the $\sigma^3$--representation becomes 
a linear operator acting on functions $f(\eta)$ and is given by
\begin{equation}\label{c.a}
H f (\eta) = - \sum_{x=1}^\ell (\eta_x - \eta_{x+1})^2  f
(\eta^{xx+1}) - \Delta \sum_{x=1}^\ell  \eta_x  \eta_{x+1}
f  (\eta)\, ,
\end{equation}
where the periodic boundary condition $\eta_{\ell+1} = \eta_1$ is
understood. $\eta^{xx+1}$ stands for the configuration $\eta$ with
occupations at sites $x$ and $x+1$ interchanged.  
The path
integral is generated by the transfer matrix
$(e^{-tH})_{\eta \eta^\prime} ~ ,~ t\geq 0  .$ For an isolated step
the first term in (\ref{c.a}) yields a symmetric, time continuous random
walk with jump rate 1 to its neighbors. For several steps
$(\eta_x - \eta_{x+1})^2$
ensures the non--crossing constraint.  $- \Delta \sum_{x=1}^{\ell}
\eta_x  \eta_{x+1}$
is a potential. Thus we see that (\ref{c.a}) indeed generates a
path integral of the form (\ref{b.e}) with $x_{j}(t)$ replaced by 
$\eta_{x}(t)$ and $P_{0}$ standing for independent random walks on 
the lattice $[1,\ldots\ell]$. For later use we define
the steps in space--time by
\medskip\\
 $\eta_x (t) = 1 \, (0)$, if at ``time'' $t$ there 
is a (no) step at site $x$,\medskip \\
$x=1, \ldots, \ell$ and  $0 \leq t \leq T$. The steps have a 
statistical weight given by (\ref{b.e}). 
 Clearly the
number of steps is $N= \sum_{x=1}^\ell  \eta _x  (t)$ independent
of $t$. 
In the dynamic picture we regard $\eta_x = 1$ as a particle and $\eta_x=0$
as no particle at site $x$ . The surface steps are then the world
lines of the particles. Particles jump to their neighboring sites, but
are never created and destroyed. We will use `step' and `particle'
interchangeably.

In the crystallographic literature our model is known as
terrace--step--kink (TSK) model. It describes a high symmetry
crystal surface miscut by a small angle. Such a vicinal
surface consists of a regular array of monoatomic steps. Through
thermal activation the steps meander but they do not cross or
terminate. The slope of the vicinal surface imposes the step density
and their average orientation. The terraces are the constant
height pieces between steps and kink refers to a step corner.

The surface defined through (\ref{b.f}), with $x_{j}(t)$ replaced by 
$\eta_{x}(t)$, is tilted in the
$x$--direction with slope $\rho = N/\ell$. A complete picture emerges only
if we tilt the surface also along the $t$--direction. To do so, we
define the step current $J_{xx+1}(t)$ through the bond $(x,x+1)$.
$J_{xx+1}(t)$
is a sequence of $\delta$--functions located at those times when
a step jumps between $x$ and $x+1$. The $\delta$--function carries
a weight $+ 1$ ($-1$) if the jump is from 
$x$ to $x+1$ ($x+1$ to $x$).
The tilt along the $t$--coordinate is enforced by the additional constraint
\begin{equation}\label{c.b}
\sum_{x=1}^\ell  \int\limits_0^T  J_{xx+1} (t) dt = N \alpha T \, ,
\end{equation}
which implies that on the average each step has the drift
$\alpha$.

 If the step variables $\eta_x  (t) ,  x=1,
 \ldots, \ell , ~0 \le t \le T$, are given, then by definition
 \begin{equation}\label{c.c}
 h(x+1,t) - h(x,t) = \eta_x(t)\, .
 \end{equation}
 (\ref{c.c}) can be integrated to yield
 \begin{equation}\label{c.d}
 h(x,t) = \sum_{y=1}^x  \eta_y (0) - \int\limits_0^t  J_{xx+1}
  (s)  ds
 \end{equation}
 with an arbitrary choice for the constant of integration.

At this point it is convenient to go from the canonical to the
grand canonical prescription. The variable conjugate to the number
of particles is the chemical potential $\mu$ and we introduce
$\lambda$ as conjugate variable to the total current (\ref{c.b}).
Thereby are obtain the transfer matrix $(e^{-tH})_{ \eta \eta^\prime} $
with hamiltonian
\begin{eqnarray}\label{c.e}
Hf(\eta)&=& - \sum_{x=1}^\ell  \{(e^\lambda  \eta_x (1-
\eta_{x+1}) + e^{-\lambda}  (1- \eta_x) \eta_{x+1})
f(\eta^{xx+1})
+ \Delta  \eta_x  \eta_{x+1}  f(\eta) - \mu \eta_x
f(\eta)\}\nonumber\\
&=& H_{\ell} (\mu ,   \lambda)  f(\eta)\, .
\end{eqnarray}
The exponential of the current gives weight $e^\lambda$ to a right jump
and weight $e^{-\lambda}$ to a left jump and the exponential of the
particle number yields the potential $\mu \sum\limits_x \eta_x$.
(\ref{c.e}) is the hamiltonian of the asymmetric XXZ model.

To make $H$ symmetric we have to continue $\lambda$ analytically to
the imaginary axis, i.e. to replace $\lambda$ by $- i
\lambda ~ \mbox{with} ~ \lambda$ real. Tracing back to the Fermi
hamiltonian (\ref{b.b}) we obtain
\begin{eqnarray}\label{c.f}
H = \sum_{x=1}^\ell \{ -e^{i \lambda}  a_x^{\dagger} a_{x+1} - e^{-i
\lambda}  a_{x+1}^{\dagger}  a_x - \Delta  a_x^{\dagger} 
a_x a_{x+1}^{\dagger}
a_{x+1} + \mu  a_x^{\dagger} a_x \}\,  .
\end{eqnarray}
This means that the dispersion relation $-2 \cos k$ is replaced
by $-2 \cos (k- \lambda).$ In the low energy limit the dominant
contributions have a a total momentum $\langle N\rangle \lambda$. Thus the
(analytically continued) $\lambda$ regulates the average fermionic
current on the ring $[1, \ldots, \ell]$,  $\lambda = 0$ corresponding
to zero current.

The thermodynamics of the surface is governed by the surface
tension $\sigma ({\bf u})$ depending on the slope ${\bf u} = (\rho,
 \rho \alpha)$.  $\sigma$ is convex up. To relate it to the hamiltonian
 (\ref{c.e}) it is convenient to define the Legendre transform, $\hat\sigma, ~
 \mbox{of}~ \sigma$ by
\begin{equation}\label{c.g}
\hat\sigma ({\bf v}) = \inf\limits_{\bf u}  
(\sigma({\bf u}) - {\bf v} \cdot {\bf u})\, .
\end{equation}
$\hat\sigma$ is convex down and in terms of $H$ in (\ref{c.e}) it is
defined by
\begin{equation}\label{c.h}
\hat\sigma (\mu, \lambda) = - \lim\limits_{T, \ell \to \infty}
\frac{1}{\ell T} ~ {\rm tr} \exp[-TH_{\ell} (\mu, \lambda)]~ . 
\end{equation}
If we take first the limit $T \to \infty$, then (\ref{c.h}) becomes $E_\ell  (\mu,
 \lambda)/ \ell$
with $E_\ell$ the ground state energy of $H_\ell$ . Thus the surface
tension is simply the Legendre transform of the ground state
energy of $H_{\ell}$ per site,
\begin{equation}\label{c.i}
\hat\sigma (\mu, \lambda) = \lim\limits_{\ell \to \infty} \, 
\frac{1}{\ell}  E_\ell  (\mu,\lambda)\, .
\end{equation}

Thermodynamic fluctuation theory suggests that a small height
fluctuation, $\delta h$, relative to the average slope ${\bf u}$ has a
probability proportional to
\begin{equation}\label{c.j}
\exp  [- \frac{1}{2}  \sum_{i,j=1}^2  \sigma_{ij} ({\bf u}) (\nabla_i
\delta h) (\nabla_j \delta h)]\,  ,
\end{equation}
where $\sigma_{ij}({\bf u}) = \partial^2 \sigma({\bf u}) / \partial u_i \partial
u_j$. Thus on a large scale the height fluctuations should be
Gaussian with a covariance
\begin{equation}\label{c.k}
( \sum_{i,j=1}^2  \sigma_{ij} ({\bf u})  k_i  k_j)^{-1}
\end{equation}
in Fourier space. (\ref{c.k}) is the covariance of a free massless
bosonic field in its Euclidean version. Fluctuations of 
two-dimensional surfaces grow
logarithmically and are therefore not stationary. They become
stationary by taking an $x$--derivate, which according to (\ref{c.c})
yields the step density. In Fourier space we only have to multiply
by $k_1^2$. The thermodynamic fluctuation theory for surfaces
predicts the low energy behavior of the density fluctuations for
the world lines of (\ref{c.e}).

We find it convenient to keep the $t$--dependence and use $k$ for
the Fourier transform with respect to $x$. Then, according to
(\ref{c.k}) and (\ref{c.c}), we have
\begin{equation}\label{c.l}
\langle \eta_x (t) \eta_{x'} (0)\rangle  - \rho^2 \simeq \int d k
e^{ik(x-x^\prime)} \frac{1}{2} c |k|e^{-\gamma|k||t| - ikvt}
\end{equation}
for large $|x-x^\prime|$ and $|t|$, where 
the parameters are defined by
\begin{equation}\label{c.m}
\sigma_{11} = (\gamma^2 + v^2)/ \gamma c ~ ,~ \sigma_{12} = -
v/ \gamma c ~,~ \sigma_{22} = 1/ \gamma c\,   .
\end{equation}
The density fluctuations are Gaussian on the scale where (\ref{c.l}) is
valid.

For the expression in (\ref{c.j}) to make sense $\sigma({\bf u})$ must be
twice differentiable at ${\bf u}$ and the matrix of second derivatives
$D^2 \sigma({\bf u})\, > \,0 .$ Already for the simple model (\ref{c.e}) with
nearest neighbor step--step interaction only, this condition is not
always satisfied. $\sigma$ is known from the Bethe ansatz
\cite{SuY1Y2}. For $\Delta > -2 ~\mbox{at}~ \rho = \frac{1}{2} ,
\alpha=0$, the steps align in antiferromagnetic order. Thereby
surface fluctuations are strongly suppressed, from logarithmic order to
order one. Changing either $\rho$ or $\alpha$ destroys this
roughening transition. We refer to \cite{ADW} for the behavior
close to the transition.

On the attractive side, $\Delta > 0$, steps may bunch to give
macroscopic patches of slope ${\bf u}=(1,0)$ and of slope ${\bf u}=(0,0)$.
This phase is bordered by the stochastic line, where $-H$ of (\ref{c.e})
generates a stochastic time evolution \cite{SuY1Y2}. Clearly,
the condition is
\begin{equation}\label{c.n}
\Delta = 2 \cosh \lambda
\end{equation}
for all $\rho$. To reexpress (\ref{c.n}) in terms of $\alpha$, we need
$\alpha= \partial e / \partial \lambda$, which is not known
in closed form,  however. For small $\lambda$ from linear response we have
$\alpha=  \rho (1-\rho) 2 \sinh \lambda$, which implies that for
small $\alpha$
\begin{equation}\label{c.o}
\Delta_c = 2 + (\alpha /2 \rho  (1- \rho))^2\,   .
\end{equation}

Coming back to (\ref{c.l}) we note that the density fluctuations decay
as $\gamma |k|$ and propagate with velocity $v$. The static
$(t=0)$ covariance is $\frac{1}{2}  c |k|$ which reflects that
steps are stiffly arranged because of the
non--crossing constraint. The parameters $\gamma, v, c$ can be expressed
through the ground state energy as
\begin{eqnarray}\label{c.p}
c^2 &=& \det  D^2 \hat\sigma = (\partial^2 e / \partial
\mu^2)(\partial^2 e / \partial \lambda^2) - (\partial^2 e /
\partial \mu \partial \lambda)^2 ,\nonumber\\
\gamma &=& c/(\partial^2 e / \partial \mu^2) ~, ~ v= (\partial^2 e /
\partial \mu \partial \lambda)/(\partial^2 e / \partial \mu^2)~,
\end{eqnarray}
where we used (\ref{c.i}), (\ref{c.m}). These relations are valid in general.

Haldane \cite{H1} observed that for Luttinger fluids the
parameters in the low energy effective bosonic action are not
independent. This is easily rederived from (\ref{c.p}). In the notation
of Haldane
\begin{equation}\label{c.q}
v_s = \gamma ~, ~ v_N = \delta \mu / \delta \rho \, .
\end{equation}
The average current is $j=\partial e /\partial \lambda$ and in
linear response $j(\lambda)= j(\lambda_0)+ v_J (\lambda_0)
(\lambda - \lambda_0)$, i.e.
\begin{equation}\label{c.r}
v_J=\partial^2 e/\partial \lambda^2 \, .
\end{equation}
The Haldane relation
reads
\begin{equation}\label{c.s}
v_N v_J = v_s^2 \,  ,
\end{equation}
i.e.
\begin{equation}\label{c.t}
(\partial^2 e /\partial \mu^2)^{-1} (\partial^2 e / \partial
\lambda^2) = \gamma^2 \,  ,
\end{equation}
which is to be compared with
\begin{equation}\label{c.u}
(\partial^2 e /\partial \mu^2)^{-1} (\partial^2 e /\partial
\lambda^2) = \gamma^2 + v^2
\end{equation}
by (\ref{c.p}). Thus the Haldane relation holds at $v=0$, equivalently
at $\lambda=0$, which he used implicitely 
by setting $j(\lambda_0) = 0$ in the linear response. The parameter $K$
mentioned in the introduction is given by
\begin{equation}\label{c.v}
K=\sqrt{v_J/v_N }= (\frac{\partial^2 e}{\partial \lambda^2}~
\frac{\partial^2 e}{\partial \mu^2})^{-1/2}
\end{equation}

For $\lambda=0$, the Hamiltonian (\ref{c.e}) is symmetric. In the
dynamic picture we have a time--reversible jump process for the
particles for which detailed balance is satisfied. As in the short
range case this gives rise to an Einstein relation and the Haldane
relation can be viewed as a particular case. 

For Brownian steps, as governed by the Hamiltonian (\ref{b.a}) with
Dirichlet conditions at $\{ x_i=x_j\}$, some simplifications
compared to the general case occur. The drift (= tilt along $t$) is
enforced by the constraint
\begin{equation}\label{c.w}
\int\limits_0^T  dt \stackrel{\cdot}{x}_j(t) = \alpha T ,
\end{equation}
$j=1, \ldots, N$. Such a drift can be trivially removed by the
global change of coordinates $y_j(t) = x_j(t) - \alpha t$. In
contrast to the terrace--step--kink model, the tilting costs only
in elastic energy for each step individually. Let $E_N$ be the
ground state energy of (\ref{b.a}) and define
\begin{equation}\label{c.x}
e(\rho) = \lim_{N,\ell \to \infty, N/\ell=\rho} \frac{1}{\ell}  E_N
\, .
\end{equation}
Then the ground state energy per unit volume at drift $\alpha$ is
given by
\begin{equation}\label{c.y}
e(\rho, \alpha)=e(\rho) + \frac{1}{2} \rho \alpha^2 \, .
\end{equation}
Comparing with (\ref{c.m}) and using ${\bf u}=(\rho, \alpha  \rho)$ we
obtain $v=\alpha$, as expected, and
\begin{equation}\label{c.z}
c = \sqrt{\rho/e^{\prime\prime}(\rho)}~, \quad \gamma = \sqrt{\rho
e^{\prime\prime}(\rho)}\, .
\end{equation}
The Haldane relation is $c\gamma=\rho$~.

We conclude this section by stating a precise conjecture. For the
terrace--step--kink model we average over $\varepsilon^{-1}$ sites
with some smooth test function $f$. From the surface picture, the
$x$ and $t$
coordinates must be on equal footing. Hence, time is also scaled
as $\varepsilon^{-1}$ and we introduce the fluctuation field
\begin{equation}\label{c.aa}
\xi^ \varepsilon (f,t) = \sum_x f(\varepsilon x)(\eta_x
(\varepsilon^{-1} t) - \rho) \, .
\end{equation}
Note that we are not in the standard central limit theorem
setting. We sum over $\varepsilon^{-1}$ sites, but expect a
fluctuation of order 1 only.
In the same spirit, for the Brownian steps we define the
fluctuation field as
\begin{equation}\label{c.ab}
\xi^\varepsilon (f,t) = \sum_j f(\varepsilon x_j (\varepsilon^{-1}
t)) - \rho \int dx f(\varepsilon x) ~.\medskip
\end{equation}
{\bf Conjecture}. {\it  We consider the path measure generated by the 
hamiltonian (\ref{c.a}) in the
limit $T \to \infty$ at fixed tilt $\alpha$, cf. (\ref{c.b}),
and in the limit $\ell \to \infty$ at fixed density $\rho = N/\ell$.
This path measure governs  the fluctuation field (\ref{c.aa}). If $e(\rho, \alpha)$, resp. 
$e(\mu, \lambda)$, is
sufficiently smooth  (at least twice differentiable),
then
\begin{equation}\label{c.ac}
\lim\limits_{\varepsilon\to 0} \quad \xi^\varepsilon (f,t) = \xi
(f,t)
\end{equation}
in distribution. The limit is jointly Gaussian with covariance
\begin{equation}\label{c.ad}
\langle \xi (f,t) \xi(g,0) \rangle ~ = \int dk \frac{1}{2} c  |k|~
e^{-\gamma|k|t|}~ e^{-ikvt}~ \hat{f}(k)^{*}\hat{g}(k) \, .
\end{equation}
The parameters $c, \gamma, v$ are given by (\ref{c.p}). Correspondingly,
for fermions in the continuum with hamiltonian (\ref{b.a}) and path measure (\ref{b.e}), the
fluctuation field (\ref{c.ab}) satisfies the limit (\ref{c.ac})
with $v = \alpha$ in (\ref{c.ad}). }

\section{Hydrodynamic fluctuation theory}\label{d}
\setcounter{equation}{0}
In the limit $T \to \infty$, for fixed $\ell$, the statistics of the steps become
stationary in $t$ and are governed by a stochastic process which
by construction is Markov. To determine its generator let $\psi$ be the
ground state and $E$ the ground state  energy for $H$ of either
(\ref{b.a}) or (\ref{c.a}), which satisfy
\begin{equation}\label{d.a}
H \psi = E \psi\, .
\end{equation}
In case of (\ref{b.a}) we have $\psi > 0$ except at coinciding points
$x_j=x_i$, $i \not= j$, where $\psi(x)=0$. For (\ref{c.a})
we first have to specify the sector $\sum_{x=1}^{\ell}  \eta_x^3 =
N$. Then, for fixed $N$,  $e^{-tH}$ has a strictly positive
integral kernel and by the
Perron-Frobenius theorem the ground state $\psi$ is unique and 
$\psi > 0$. The (backwards) generator of the Markov process
for the steps is defined by
\begin{equation}\label{d.b}
L f = - \psi^{-1}  (H-E) \psi f
\end{equation}
as acting on functions $f$ over the configuration space. The Markov
process has $\psi^2$
as unique invariant measure.

We carry out this construction for the XXZ model in the $\sigma^3$
representation  and obtain, in the notation of (\ref{c.a}),
\begin{equation}\label{d.c}
L f(\eta) = \sum_{x=1}^l  (e^\lambda  \eta_x
(1- \eta_{x+1}) +  e^{- \lambda}  (1- \eta_x) \eta_{x+1})
c_{xx+1} (\eta) [f(\eta^{xx+1}) - f(\eta)] \, .
\end{equation}
$L$ is the generator for a stochastic lattice gas, where particles
jump to their neighboring sites. The exchange rate, $c_{xx+1}(\eta),$
between sites $x ~ \mbox{and}~ x+1$ is given by
\begin{equation}\label{d.d}
c_{xx+1} (\eta)= \psi (\eta^{xx+1})/ \psi(\eta) \,.
\end{equation}
The jumps to the right are biased by the factor $e^\lambda$ and
those to the left by $e^{-\lambda}$. If $\lambda=0$, $H$ is
symmetric and the stochastic evolution satisfies detailed balance.

The rates (\ref{d.d}) are determined through the ground state, which is not
known in general. Only for $\Delta=0$ one has the explicit ground
state $\psi$.
If we denote by $x_1, \ldots, x_N$ the positions of the particles
in the sector $\sum_{x=1}^\ell  \eta_x=N,$ then in this
sector
\begin{eqnarray}\label{d.e}
\psi(x_1,\ldots, x_N) = \prod_{i<j=1}^N | \sin  (\pi
(x_i-x_j) /\ell) | \,.
\end{eqnarray}
Therefore the exchange rates are given by
\begin{equation}\label{d.f}
c_{xx+1}^{(N)} = \exp [- (\eta_{x+1} - \eta_x)
\sum_{y=1\atop y \not= x,x+1}^\ell (\log  (\frac{\sin
(\pi(y-x+1)/\ell)}{\sin (\pi(y-x)/\ell)})) \eta_y ] \, .
\end{equation}
Taking formally the infinite volume limit yields
\begin{equation}\label{d.g}
c_{xx+1} = \exp [- (\eta_{x+1} - \eta_x)
\sum_{y\not= x,x+1} (\log  (\frac{y-x+1}{y-x}) )\eta_y ] \,.
\end{equation}

(\ref{d.g}) teaches us several points. The rates are repulsive: more
particles to the left of $x$ favors a right jump of the particle
at $x$. The  rates are long ranged, which means that the finite
range intuition is no longer applicable. Since $\log ((y-x-1)/y-x) \cong
-(y-x)^{-1}$, at infinite volume the rates may be infinite. The
dynamics with rates (\ref{d.g}) is then defined only for $| \psi|^2$
almost all configurations. Because the case $\Delta=0$ maps to a free
fermion theory, one can construct the infinite volume ground state
as a measure on particle configurations and the Markov semigroup $e^{Lt}$
in the corresponding $L^2$--space \cite{Sp1,PrSp2}. This implies
the almost sure existence of the dynamics. If one adds the
nearest neighbor interaction $\Delta$, one expects exchange rates to 
be  
qualitatively
similar to (\ref{d.g}). We are not aware of any result in
this direction.

For the Brownian steps $L$ is the generator of a diffusion process
given by
\begin{equation}\label{d.h}
L f(x) = \sum_{j=1}^N  (a_j(x)
\frac{\partial}{\partial x_j} + \frac{\partial^2}{\partial x_j^2})
 f(x)\, ,
\end{equation}
$x= (x_1, \ldots, x_N)$. The drift on the $j$--th particle is
\begin{equation}\label{d.i}
a_j = - \frac{\partial}{\partial x_j} \log \psi \, .
\end{equation}
Thus $\log \psi$ is the potential for the diffusion process.

For Brownian steps with exclusion only, i.e. $V=0$, one finds that
\begin{equation}\label{d.j}
a_j(x) = \sum_{i=1,i\not=j}^N  \cot (\pi(x_j-x_i)/\ell)\, .
\end{equation}
For $\ell\to \infty$ this expression converges to
\begin{equation}\label{d.k}
a_j(x) = \sum_{i,i\not=j} \frac{1}{x_j-x_i} \, ,
\end{equation}
i.e. to a repulsive $1/x$--force as for the lattice gas. The model
with drift
(\ref{d.k}) was introduced by Dyson \cite{Dy1}. Its potential is formally
given by
\begin{equation}\label{d.l}
- \sum_{i\not=j}  \log |x_i-x_j| \, .
\end{equation}
In fact, Dyson adds a confining external potential, which is
quadratic and considers (\ref{d.k}) for finite $N$.

We conclude that density fluctuations in the fermionic ground state may
equally well be studied through the stochastic dynamics (\ref{d.c}),
(\ref{d.h}). If the potential is short range, there is a well developed
machinery known as hydrodynamic fluctuation theory. One argues
(and proves in many model systems \cite{Sp3,Yau1}) that density
fluctuations relax diffusively and are driven by a white noise
random flux. This corresponds to the linear Langevin equation
\begin{equation}\label{d.m}
\frac{\partial}{\partial t} \xi(x,t) = D \frac{\partial^2}{\partial x^2}
 \xi(x,t)+ \sqrt{\sigma} \frac{\partial}{\partial x}  W(x,t)\, .
\end{equation}
$D$ is the bulk diffusion coefficient and $\sigma$ is the  bulk
mobility. They are related through $\sigma = \chi D ~ \mbox{with}~ \chi$
the static compressibility of the lattice gas. $W(x,t)$ is
space--time white noise. The spatial derivative ensures the
conservation of the number of particles. The stationary measure
for (\ref{d.m}) is Gaussian with covariance matrix $(\sigma/2D)$. In
passing we mention that in the short range case (\ref{d.m}) holds only
for $\lambda=0$, i.e. in the symmetric case. For $\lambda \not=0$
one switches over to the KPZ universality class. It is
characterized by the dynamic exponent $z=3/2$, rather than $z=2$
as in (\ref{d.m}). The fluctuations are non--Gaussian \cite{J1}.

Returning to the long range case of interest here, we have already
obtained the limiting Gaussian fluctuations in the Conjecture. We
write $\xi(f,t) = \int dx  f(x)\xi(x,t)$ and note that (\ref{c.ad}) 
determines
a semigroup in $t$. Therefore (\ref{c.ad}) must be the (stationary)
solution of the linear Langevin equation
\begin{equation}\label{d.n}
\frac{\partial}{\partial t}  \xi(x,t) = (- \gamma \sqrt{- \partial^2/
\partial x^2} - v \frac{\partial}{\partial x})  \xi(x,t) +
\sqrt{c \gamma} \frac{\partial}{\partial x}  W(x,t)\, ,
\end{equation}
which is a surprisingly minimal modification compared to the short
range case (\ref{d.m}). $\gamma$ plays the role of $D$ and $c$ the one
of $\sigma$. The crucial difference is that a Fourier mode $e^{ikx}$
decays as $e^{-\gamma |k||t|}$ rather than as $e^{-Dk^2|t|}$.

\section{Scaling limit}\label{e}
\setcounter{equation}{0}
We support our general conjecture by arguing that the
stochastic particle evolution is close to the Langevin equation 
(\ref{d.n}). We
will do so on a fairly formal level. In particular, we simply work
in infinite volume. In the appendix we explain how parts of our
arguments can be made rigorous.

Our strategy is most easily explained for the Brownian step model.
The equations of motion are
\begin{equation}\label{e.a}
\frac{d}{dt} x_j(t) = a_j(x(t)) + \dot{b}_j(t) \, ,
\end{equation}
where $\dot{b}_j(t)$ is normalized white noise
independent for each $j$. The scaled fluctuation field satisfies
then the differential equations
\begin{equation}\label{e.b}
\frac{d}{dt}  \xi^\varepsilon (f,t) = \sum\limits_j  f^\prime
(x_j^\varepsilon(t))
a_j (x^\varepsilon(t)) + \varepsilon \sum\limits_j  
f^{\prime\prime}(x_j^\varepsilon(t))
 + \sqrt{\varepsilon} \sum\limits_j  f^\prime
 (x_j^\varepsilon(t)) \dot{b}_j(t) \,.
\end{equation}
Here $x_j^\varepsilon(t) = \varepsilon  x_j(\varepsilon^{-1} t)$
and we used the scale invariance of white noise as
$\dot{b}_j(\varepsilon^{-1}t) = \sqrt{\varepsilon}  
\dot{b}_j(t)$.

The second term in (\ref{e.b}) converges to $\rho \int dx f^{\prime\prime}(x) =
0$. The third term converges to a space--time Gaussian measure
with covariance
\begin{equation}\label{e.c}
\delta (t-t^\prime)  \rho \int dx  f^\prime(x) g^\prime (x) \, .
\end{equation}
This uses only that with respect to the distribution given by $|\psi|^2$
we have $\varepsilon  \sum\limits_j  f(\varepsilon x_j) 
\to \rho \int dx f(x)$
in probability as $\varepsilon \to 0.$ (\ref{e.c}) is in accordance
with (\ref{d.n}), since $c \gamma = \rho$. Thus we are ``only''  left
with to show that (recall that $v=0$)
\begin{equation}\label{e.d}
\int\limits_0^t ds  \sum\limits_j  f^\prime (x_j^\varepsilon (s))
a_j(x^\varepsilon(s))  ds
\simeq - \int\limits_0^t \sum\limits_j \gamma
\sqrt{-\partial^2/\partial x^2}  f  (x_j^\varepsilon(s))  ds \, .
\end{equation}
If so, the integrated version of (\ref{e.b}) becomes a closed equation
and agrees with (\ref{d.n}).

To establish (\ref{e.d}) is certainly the hard part of the matter. In
Appendix A we prove that, if we assume the 2-point function to
scale, the substitution (\ref{e.d}) holds. This is of interest
because the scaling of the 2--point function by itself implies
already that the fluctuations are Gaussian. I would not know how
to obtain such a result otherwise.

A further example for wich the substitution (\ref{e.d}) holds is the
Calogero--Sutherland  model \cite{Sp4}, where the pair potential
is proportional to $1/x^2$. If the particles are on a ring
$[0,\ell]$ and if we take all the image potentials into account,
then the ground state wave function of the Calogero--Sutherland
model is
\begin{equation}\label{e.e}
\psi(x_1, \ldots, x_N)= \prod_{i<j=1}^N | \sin
(\pi(x_i - x_j) /\ell)|^{\beta/2}\, ,
\end{equation}
 $\beta > 0$. By (\ref{d.i}) the corresponding drift is given by
 \begin{equation}\label{e.f}
 a_j (x_1, \ldots, x_N) = \frac{\beta}{2}
 \sum_{i=1,i \not=j}^N \frac{\pi}{\ell} \cot (\pi
 (x_i-x_j)/\ell)
 \end{equation}
 which in the limit $\ell \to \infty$ yields
 \begin{equation}\label{e.g}
 a_j (x) = \sum_{i=1,i \not=j}^N \frac{\beta}{2}
  \frac{1}{x_j-x_i} \,.
 \end{equation}
 Compared to (\ref{d.h}) only the strength is changed.

The ground state energy per length is $e(\rho)=\frac{1}{24}  
\beta^2 \pi^2 \rho^3$
(note that our kinetic energy is $-\frac{1}{2} \Delta)$.
Therefore, by (\ref{c.z}),
\begin{equation}\label{e.h}
c=\frac{2}{\pi \beta} \, , \quad \gamma = \frac{1}{2} \pi
\beta \rho \, .
\end{equation}
For the Calogero--Sutherland model the substitution in (\ref{e.d}) holds
without averaging in time. This is somewhat surprising and very
particular for the $1/x^2$ potential. To complete the argument one
needs the central limit result of Johansson \cite{J},
\begin{equation}\label{e.i}
\lim_{\varepsilon \to 0}\,  \langle \exp  [ \sum_j  f(\varepsilon x_j)]
\rangle =  \exp
 [ \frac{1}{2}  \frac{1}{\pi \beta} \int dk |k||
\hat{f}_k|^2]
\end{equation}
for smooth test functions with $\int dx f(x) = 0$, where $\langle \cdot
\rangle$
is the average in the ground state (\ref{e.e}) in the infinite volume
limit.
The instructive computation is explained in Appendix B.

Turning to the terrace--step--kink model the situation is more
complicated, as may be anticipated from the short range exchange
rates \cite{Yau2V}. We consider the symmetric case and set
$\lambda=0$. Using the method of martingales one can show that the
variance of the time--integrated noise is given by
\begin{equation}\label{e.j}
\langle c_{01} (\eta) \rangle t = (\sum_\eta \psi(\eta)
\psi(\eta^{01})(\eta_0-\eta_1)^2) t \, ,
\end{equation}
which is to be compared with the prediction
\begin{equation}\label{e.k}
c\gamma = \frac{\partial^2 e}{\partial \lambda^2}
\mid_{\lambda=0}
\end{equation}
from (\ref{d.n}). Note that  $v=0$, since $\lambda=0$. By second order
perturbation theory in $\lambda$ we obtain
\begin{eqnarray}\label{e.l}
\frac{\partial^2 e}{\partial \lambda^2} &=& \sum_\eta \psi(\eta)
\psi(\eta^{01})(\eta_0-\eta_1)^2\nonumber\\
&-& \sum_x\sum_\eta   \psi(\eta^{01})
(\eta_0- \eta_1)(H-E)^{-1}  (\eta_x-\eta_{x+1})
\psi(\eta^{xx+1}) \, .
\end{eqnarray}
Thus (\ref{e.k}) holds only if the second term vanishes.

For $\Delta=0$ the sum $\sum\limits_x  (\eta_x- \eta_{x+1})  \psi(\eta^{xx+1})$
is the total current $J$ acting on $\psi$. Since $J\psi = 0$, we
conclude that for free fermions (\ref{e.k}) holds. In this case
the two--point is explicitely known and we can use the argument of
Appendix A to prove the Conjecture. Alternatively the $n$--point
density correlations can be written in terms of 2--point
functions. By applying the closed loop theorem one again concludes
that the scaling limit is Gaussian with covariance (\ref{c.ad})
\cite{PrSp2,M1}.

If $\Delta \not=0$, the second term in (\ref{e.l}) is not expected to
vanish. This can be verified by expanding $(H-E)^{-1}$ to second
order in $\Delta$. Thus the martingale term (\ref{e.j}) yields the
wrong noise strength in the Langevin equation - the hallmark of
the so--called non--gradient systems. The drift term
\begin{equation}\label{e.m}
\int\limits_0^t ds \sum_x  f^\prime (\varepsilon x)  c_{xx+1}
(\eta(\varepsilon^{-1}s))
\end{equation}
can no longer substituted deterministically as in (\ref{e.d}). Somewhere
hidden there must be an extra noise term. While for short range
lattice gases this mechanism is understood by the beautiful work
of Varadhan and Yau \cite{Yau2V}, cf. also \cite{KiLa}, the
situation looks rather complicated for the long range case
considered here.

\section{Conclusions}\label{f}
\setcounter{equation}{0}
We tied together three, at first sight, disconnected pieces: the
Luttinger liquid behavior at low energy, surface fluctuations, and
the hydrodynamic fluctuation theory for the stochastic dynamics of
world lines.

In the surface picture one can easily identify the universal
low energy limit
of the density fluctuations for one-dimensional Fermi
fluids. Of course, as an input one needs a variant of the Einstein
fluctuation formula. The remainder of the argument is then
straightforward. In particular, we show that the parameters of the
Tomanaga--Luttinger Hamiltonian must be matched to suitable
second derivatives of the energy/length, c.f. Eqs. (\ref{c.p}). To our
knowledge, this has not been discussed with sufficient clarity
before.

In the interpretation of the world lines as stochastic dynamics 
the particles have long
range interactions, which result from the Dirichlet boundary
condition at coinciding positions and thus from the Pauli
exclusion principle. We develop a novel fluctuation theory for
the long range case. In fact, the only modification is to replace
in the drift term the Laplacian,  $-\partial^2/\partial x^2$, by
the nonlocal integral operator $\sqrt{-\partial^2 / \partial x^2}.$
Heuristically, following the arguments of Tomanaga, the $|k|$
decay comes from linearizing the dispersion relation of the
particles at the Fermi surface. The not so obvious point is that an
interaction changes only the prefactor $\gamma$ but not the decay
law itself. At
least for the Calogero--Sutherland hamiltonian, the mechanism
behind such a renormalization could be fully elucidated.

\begin{appendix}
\renewcommand{\theequation}{\thesection.\arabic{equation}}
\section{}\label{g}
\setcounter{equation}{0}
For the Brownian steps we define the structure function, $S(k,t)$, by
\begin{equation}\label{g.a}
\langle (\sum_j  f(x_j(t))) (\sum_j g(x_j)) \rangle  =  \int dk
\hat{ f}(k)^{*}g(k) S(k,t)
\end{equation}
with  $\int dx f(x) = 0 = \int dxg(x)$.
The average is in the stationary process at infinite volume with
density $\rho$. The existence of this limit and its spatial
ergodicity is assumed here. In principle, this could be avoided by
considering a finite circle of length $\ell$, for which the
stationary process exists, and by scaling $\ell=\varepsilon^{-1} \rho^{-1},
N= \varepsilon^{-1}$. We adopt this strategy for the
Calogero--Sutherland model in Appendix B.
In essence, Fourier space becomes then discrete, $k \in 2 \pi \mathbb Z $.

Our real assumption is the scaling of the structure function as
\begin{equation}\label{g.b}
\lim_{\varepsilon \to 0}  \varepsilon^{-1} S(\varepsilon k,
\varepsilon^{-1} t)= \frac{c}{2} |k|  e^{-\gamma|k||t|}
\end{equation}
with $c=\sqrt{\rho / e^{\prime\prime}}, ~ \gamma = \sqrt{\rho
e^{\prime\prime}}$. We want to prove that (\ref{g.b}) implies
\begin{equation}\label{g.c}
\lim_{\varepsilon \to 0}\langle[\int\limits_0^t ds \sum_j 
 a_j(x^\varepsilon(s))
 \ f^\prime(x_j^\varepsilon (s)) + \int\limits_0^t ds \sum_j
 \gamma \sqrt{-\partial^2/\partial x^2}  f(x_j^\varepsilon(s))]^2  \rangle = 
 0\, .
\end{equation}
We note that the time average is needed. This reflects
that the system takes a while to adjust to local perturbations.

We work out the square and use that $L$ is symmetric,
 $\langle F(LG) \rangle = \langle G(LF) \rangle$,
and
\begin{eqnarray}\label{g.d}
&&j(f^\prime)(x) = \sum_j a_j(x)  f^\prime(x_j) + \sum_j
\frac{1}{2}  f^{\prime \prime}(x_j) = L n(f)(x)\, , \nonumber\\     
&&n(f)(x)=\sum_j  f(x_j)\,   .
\end{eqnarray}
Under our scaling the term $n(f^{\prime\prime})$ vanishes. We have
\begin{eqnarray}\label{g.e}
&&\int\limits_0^t ds_1 \int\limits_0^t ds_2  \langle 
j^\varepsilon(f^\prime, s_1) j^\varepsilon (f^\prime, s_2) \rangle 
\nonumber\\
&& = \int\limits_0^t ds_1 ( \int\limits_0^{s_1} ds_2  \langle
j^\varepsilon(f^\prime)  e^{L(s_1-s_2)\varepsilon^{-1}}  j^\varepsilon
(f^\prime) \rangle 
+ \int\limits_{s_1}^t ds_2  \langle j^\varepsilon(f^\prime)
e^{L(s_2-s_1) \varepsilon^{-1}} j^\varepsilon(f^\prime)  \rangle 
) \nonumber\\
&&=\int\limits_0^t ds_1 (-\int\limits_0^{s_1} ds_2
\frac{\partial}{\partial s_2}  \langle j^\varepsilon(f^\prime)  
e^{L(s_1-s_2)\varepsilon^{-1}}
 n^\varepsilon (f)  \rangle  
  +\int\limits_{s_1}^t ds_2  
 \frac{\partial}{\partial s_2}  \langle j^\varepsilon(f^\prime)
e^{L(s_2-s_1)\varepsilon^{-1}} n^\varepsilon(f)  \rangle) \nonumber\\
&&= -2t  \langle j^\varepsilon(f^\prime)n^\varepsilon(f)   \rangle + 2
\int\limits_0^t ds \frac{\partial}{\partial s}  \langle
n^\varepsilon(f) e^{Ls \varepsilon^{-1}} n^\varepsilon (f)  \rangle 
\nonumber\\
&&= -2t  \langle j^\varepsilon(f^\prime)n^\varepsilon(f)  \rangle + 2
\langle n^\varepsilon(f,t) n^\varepsilon (f)  \rangle
-2  \langle n^\varepsilon(f) n^\varepsilon(f) \rangle \, .
\end{eqnarray}
For the first summand we have
\begin{equation}\label{g.f}
\langle j^\varepsilon(f^\prime) n^\varepsilon(f)  \rangle  = -\varepsilon^{-1}
\frac{1}{2}  \sum_j  \langle \frac{\partial}{\partial x_j}
n^\varepsilon(f) \frac{\partial}{\partial x_j} n^\varepsilon(f) 
\rangle \, .
\end{equation}
Therefore in the limit $\varepsilon \to 0$ we obtain
\begin{equation}\label{g.g}
\int dk | \hat f|^2 (t|k|^2 \rho - c|k| (1 - e^{-\gamma|k||t|})) \, .
\end{equation}

The second term reads, with $g=-\gamma \sqrt{-\partial^2 /\partial
x^2} f,$
\begin{eqnarray}\label{g.h}
& & -2 \int\limits_0^t ds_1 \int\limits_0^t ds_2 \langle n^\varepsilon (g,
s_1) j^\varepsilon (f^\prime, s_2) \rangle\nonumber\\
&=& -2\int\limits_0^t ds_1 [\int\limits_0^{s_1} ds_2 \langle n^\varepsilon
(g)  e^{L(s_1-s_2) \varepsilon^{-1}}
j^\varepsilon (f^\prime) \rangle
 + \int\limits_{s_1}^t ds_2 \langle j^\varepsilon
(f^\prime) e^{L(s_2-s_1) \varepsilon^{-1}} n^\varepsilon (g)
\rangle]\nonumber\\
&=& -2\int\limits_0^t ds_1 [-\int\limits_0^{s_1} ds_2
\frac{\partial}{\partial s_2} \langle n^\varepsilon
(g)  e^{L(s_1-s_2) \varepsilon^{-1}} n^\varepsilon (f)
\rangle
+ \int\limits_{s_1}^t ds_2
\frac{\partial}{\partial s_2} \langle j^\varepsilon
(f^\prime) e^{L(s_2-s_1) \varepsilon^{-1}} n^\varepsilon (g)
\rangle]\nonumber\\
&=& -2 \int\limits_0^t ds [ - 2 \langle n^\varepsilon
(g) n^\varepsilon(f) \rangle +2 \langle n^\varepsilon (g)
 e^{L s\varepsilon^{-1}} n^\varepsilon(f) \rangle ]\, .
\end{eqnarray}
By assumption this expression converges to
\begin{equation}\label{g.i}
\int dk |\hat f|^2  (-2 t \gamma c k^2 + 2 |k| c (1 - 
e^{-\gamma|k|t}))\,  .
\end{equation}

The third term is given by
\begin{equation}\label{g.j}
\int\limits_0^t ds_1\int\limits_0^t ds_2 \langle n^\varepsilon (g,s_1)
n^\varepsilon (g,s_2) \rangle \,,
\end{equation}
which by assumption converges to
\begin{equation}\label{g.k}
\int\limits_0^t ds_1 \int\limits_0^t ds_2  \int dk |\hat f|^2 
\gamma^2 \frac{c}{2}
|k|^3 e^{-\gamma |k||s_1-s_2|}\nonumber\\
= \int dk |\hat f|^2  (t \gamma c k^2 - c |k| (1 - 
e^{-\gamma|k|t}))\, .
\end{equation}
Using that $\gamma c= \rho$, the sum of the three terms vanishes, as
claimed.

To complete the argument one notes that
\begin{equation}\label{g.l}
M^\varepsilon (f,t)=\int\limits_0^t ds \sum_j f(\varepsilon x_j
(\varepsilon^{-1} s)) d b_j(s)
\end{equation}
is a martingale with square
\begin{equation}\label{g.m}
M(f,t)^2 = \int\limits_0^t ds \varepsilon \sum_j f^\prime
(\varepsilon x_j (\varepsilon^{-1} s))^2 + M_1^\varepsilon (f,t)\, ,
\end{equation}
where $M_1^\varepsilon (f,t)$ is again a martingale. We assume now
the validity of a law of large numbers as
\begin{equation}\label{g.n}
\lim_{\varepsilon\to 0}  \varepsilon \sum_j g (\varepsilon x_j)
= \rho \int dx g(x) \, .
\end{equation}
Then, under suitable tightness, in the limit $\varepsilon\to 0$
\begin{equation}\label{g.o}
M(f,t)^2 = t \int dx f^\prime(x)^2 + M_1(f,t)
\end{equation}
with both $M(f,t)$ and $M_1(f,t)$ martingales. This implies that $M(f,t)$
is Brownian motion in $t$ with covariance $\int dx f^\prime
(x)^2$. We refer to \cite{Sp4} for a more complete discussion.

\section{}\label{h}
\setcounter{equation}{0}
We consider the Calogero--Sutherland model on a finite ring of
length $\ell$. The density is $\rho = N/\ell$. We scale the positions
with $\varepsilon^{-1}$. Setting $\varepsilon = 2 \pi / \ell$, we
scale back to the circle $[0,2 \pi]$. Then $N =\varepsilon^{-1} 2 \pi \rho
$.
Since the density scales, we set $ 2 \pi \rho = 1$
and use $N $ instead of $\varepsilon$. After these
transformations the ground state $\psi^2 = e^{-\beta V}$
  with the logarithmic potential
\begin{equation}\label{h.a}
V= - \sum_{i<j=1} \log |\sin (x_i - x_j) /2)|\, .
\end{equation}
The drift is then
\begin{equation}\label{h.b}
a_j(x) = - \frac{\partial}{\partial x_j} \beta V(x) =
\frac{\beta}{2} \sum_{i=1\atop_{i \not= j}}^N \cot ( (x_j - x_i)/2) \, .
\end{equation}
For this model we  show the
validity of the substitution (\ref{e.d}) 
without averaging in $t$.

More explicitely, we have to show that
\begin{equation}\label{h.c}
\frac{1}{N} \sum_{j=1}^N a_j(x) f^\prime(x_j) = \frac{1}{2 N}
\sum_{i,j=1\atop_{i \not=j}}^N \beta \cot ((x_j - x_i)/2)
f^\prime(x_j)\nonumber\\
\simeq \sum_j g(x_j)\, ,
\end{equation}
where
\begin{equation}\label{h.d}
g(x) = \frac{\beta}{4 \pi} \int\limits_0^{2 \pi} dy \cot ((x-y)/2)
f^\prime(y)\, ,
\end{equation}
in the sense of the Cauchy principal part. Since we are on a ring,
$f$ has discrete Fourier coefficients $\hat f_k = (1/2 \pi)
\int_0^{2\pi} dx  e^{ikx} f(x)$.
We assume that 
$\sum_{|k|}
|k|^3 \hat f_k < \infty$. Note that $\hat g_k = -(\beta /2)|k| \hat
f_k$. To show (\ref{h.c}) we use a result of Johansson \cite{J} which
states
\begin{equation}\label{h.e}
\lim_{N\to\infty} \frac{1}{N} \langle \sum_{j=1}^N h(x_j) \rangle = \hat h_0
\end{equation}
and, provided $\hat h_0 = 0 = \hat f_0$,
\begin{equation}\label{h.f}
\lim_{N\to\infty} \langle (\sum_{j=1}^N  h(x_j)) (\sum_{i=1}^N f(x_i))
\rangle = \frac{2}{\beta} \sum_k |k| \hat h_k^* \hat f_k \, .
\end{equation}

Let us work out the square. The first term is
\begin{equation}\label{h.g}
\langle (\sum_{j=1}^N g(x_j))^2 \rangle\, ,
\end{equation}
which by (\ref{h.f}) tends in the limit $N\to\infty$ to
\begin{equation}\label{h.h}
\frac{2}{\beta} \sum_k |k| |\hat g_k|^2 = \frac{\beta}{2} \sum_k
|k|^3 |\hat f_k|^2\, .
\end{equation}

For the second term we use $\frac{\partial}{\partial x_i}  e^{-\beta V} =
a_i  e^{-\beta V}$. Then
\begin{eqnarray}\label{h.i}
& &-2  \frac{1}{N} \langle (\sum_{j=1}^N a_j f^\prime(x_j))(\sum_{i=1}^N
g(x_i)) \rangle = \frac{2}{N}  \sum_{i,j=1}^N \langle
\frac{\partial}{\partial x_j} (f^\prime(x_j) g (x_i))
\rangle\nonumber\\
& & = \frac{2}{N} \langle (\sum_{j=1}^N f^{\prime\prime}(x_j))(\sum_{i=1}^N
g(x_i)) \rangle + \frac{2}{N} \langle \sum_j f^\prime(x_j) g^\prime(x_j)
\rangle\, .
\end{eqnarray}
By (\ref{h.f}) the first summand vanishes and by (\ref{h.e}) the second
summand converges to
\begin{equation}\label{h.j}
2\sum_k |k|^2 \hat f_k^{*} \hat g_k = - 2 \beta \sum_k |k|^3 |\hat
f_k|^2 \, .
\end{equation}
For term number three we use that\\ $(\partial^2/\partial x_i \partial
x_j) e^{-\beta V} = a_i a_j e^{-\beta V} + (\partial a_i/ \partial x_j)
e^{-\beta
V}$. Then
\begin{eqnarray}\label{h.k}
& &\frac{1}{N^2} \langle (\sum_{j=1}^N a_j f^\prime(x_j))(\sum_{i=1}^N
a_i f^\prime(x_i) )\rangle \nonumber\\
&=&\frac{1}{N^2} \sum_{i,j=1}^N  \langle \frac{\partial}{\partial x_i}
\frac{\partial} {\partial_{xj}}(f^\prime(x_i)f^\prime(x_j)) -
(\frac{\partial }{\partial x_j} a_i)
f^\prime(x_i)f^\prime(x_j)\rangle \nonumber\\
&=&\frac{1}{N^2} ( \langle (\sum_{j=1}^N  f^{\prime\prime}(x_j))(\sum_{j=1}^N
f^{\prime\prime}(x_i)) \rangle + \langle \sum_{j=1}^N f^{\prime\prime}(x_j)^2
\rangle 
+ 2 \langle \sum_j f^\prime(x_j) f^{\prime\prime\prime}(x_j)\rangle
\nonumber\\
&+&\frac{1}{N^2}  \frac{\beta}{8}\langle \sum_{i,j=1}^N |\sin((x_i-x_j)/2)
| ^{-2} [f^\prime(x_i)-  f^\prime(x_j)]^2\rangle\, .
\end{eqnarray}
By (\ref{h.e}) the first summand vanishes. The second summand converges
to
\begin{equation}\label{h.l}
 \frac{\beta}{8} (2 \pi)^{-2} \int\limits_0^{2 \pi} dx \int\limits_0^{2 \pi}
dy |\sin((x-y)/2)|^{-2} (f^\prime(x) - f^\prime(y))^2 
=\frac{\beta}{2} \sum_k |k|^3 |\hat f_k |^2\,.
\end{equation}
Adding the three terms we conclude that the sum vanishes.
\bigskip\\
\end{appendix}
{\bf Acknowledgements}: I thank M. Pr\"{a}hofer for most helpful 
discussions.

\end{document}